\documentclass[pra,twocolumn,superscriptaddress,showpacs,floatfix]{revtex4}
\usepackage{graphicx}
\usepackage{amsmath}
\usepackage{amssymb}

\newcommand{\rem}[1]{}

\begin{document}

\title{Computing the distance between quantum channels: \\
Usefulness of the Fano representation}
\author{Giuliano Benenti}
\email{giuliano.benenti@uninsubria.it}
\affiliation{CNISM, CNR-INFM, and Center for Nonlinear and Complex Systems,
Universit\`a degli Studi dell'Insubria, via Valleggio 11, 22100 Como, Italy}
\affiliation{Istituto Nazionale di Fisica Nucleare, Sezione di Milano,
via Celoria 16, 20133 Milano, Italy}
\author{Giuliano Strini}
\email{giuliano.strini@mi.infn.it}
\affiliation{Dipartimento di Fisica, Universit\`a degli Studi di Milano,
via Celoria 16, 20133 Milano, Italy}
\date{\today}
\begin{abstract}
The diamond norm measures the distance between two quantum channels.
From an operational vewpoint, this norm measures how well we can 
distinguish between two channels by applying them to input states 
of arbitrarily large dimensions. In this paper, we show that the 
diamond norm can be conveniently and in a physically transparent way 
computed by means of a Monte-Carlo algorithm based on the Fano 
representation of quantum states and quantum operations.  
The effectiveness of this algorithm is illustrated for several 
single-qubit quantum channels.
\end{abstract}

\pacs{03.65.Yz, 03.67.-a}

\maketitle

\section{introduction}
\label{sec:introduction}

Quantum information processes in a noisy environment 
are conveniently described in terms of
\textit{quantum channels}, that is, linear, trace preserving, 
completely positive maps on the set of quantum states~\cite{qcbook,nielsen}.
The problem of discriminating quantum channels is of great interest. 
For instance, knowing the correct noise model might provide useful
information to devise efficient error-correcting strategies, both 
in the fields of quantum communication and quantum computation. 

It is therefore natural to consider \textit{distances} between quantum 
channels, that is to say, we would like to quantify how similarly
two channels $\mathcal{E}_1$ and $\mathcal{E}_2$
act on quantum states, or in other words to determine
if there are input states $\rho$ on which the two channels 
produce output states $\rho_1^\prime=\mathcal{E}_1(\rho)$ and 
$\rho_2^\prime=\mathcal{E}_2(\rho)$ that are distinguishable. 
The trace norm of $\rho_1^\prime-\rho_2^\prime$ represents 
how well $\rho_1^\prime$ and $\rho_2^\prime$ 
can be distinguished by a measurement~\cite{helstrom}:
the more orthogonal two quantum states are, the easier it is to 
discriminate them. The trace distance of two quantum channels is
then obtained after maximizing the trace norm 
of $\rho_1^\prime-\rho_2^\prime$ over the input state $\rho$.
However, the trace norm is not a good measure of
the distance between quantum channels. 
Indeed, in general the presence in the input state 
of entanglement with an ancillary system can help 
discriminating quantum channels~\cite{preskill,acin,dariano,sacchi,sacchi2}. 
This fact is captured by the \textit{diamond norm}~\cite{kitaev,watrous}:
the trace distance between the overall output states (including 
the ancillary system) is optimized over all possible input states,
including those entangled.

The computation of the diamond norm is not known to be straightforward
and only a limited number of algorithms have been 
proposed~\cite{JKP09,watrous09,BT10}, based on complicated 
semidefinite programming or convex optimization.
On the other hand, analytical solutions are limited to 
special classes of channels~\cite{sacchi,sacchi2}.
In this paper, we propose a simple and 
easily parallelizable Monte-Carlo algorithm based on the 
Fano representation of quantum states and quantum operations. 
We show that our algorithm provides reliable results for the 
case, most significant for present-day implementations of 
quantum information processing, of single-qubit quantum channels. 
Furthermore, in the Fano representation quantum operations are
described by affine maps whose matrix
elements have precise physical meaning: 
They are directly related to the evolution of 
the expectation values of the system's
polarization measurements~\cite{qcbook,BFS,BS09}.

The paper is organized as follows. After reviewing in Sec.~\ref{sec:dnorm} 
basic definitions of the distance between quantum channels,
we discuss in Sec.~\ref{sec:methods}
two numerical Monte-Carlo strategies for computing the diamond
norm. The first one is based on the Kitaev's characterization
of the diamond norm.
The second one is based on the Fano representation of quantum states
and quantum operations. The two methods are compared in 
Sec.~\ref{sec:1qchannels} for a few physically significant single-qubit
quantum channels. Finally, our conclusions are drawn in Sec.~\ref{sec:conclusions}.

\section{The diamond norm}
\label{sec:dnorm}

\subsection{Basic definitions}
\label{subsec:basics}

We consider the following problem: given two quantum channels 
$\mathcal{E}_1$ and $\mathcal{E}_2$, and a single channel use, 
chosen uniformly at random from $\{\mathcal{E}_1,\mathcal{E}_2\}$, 
we wish to maximize the probability of correctly identifying
the quantum channel. It seems natural to reformulate the optimization
problem into the problem of finding an input state (density matrix)
$\rho$ in the Hilbert space $\mathcal{H}$ such that the 
error probability in the discrimination of the output states 
$\mathcal{E}_1(\rho)$ and $\mathcal{E}_2(\rho)$ is minimal.
In this case, the minimal error probability reads 
\begin{equation}
\begin{array}{c}
{\displaystyle
p_E'= \frac{1}{2}-\frac{||\mathcal{E}_1-\mathcal{E}_2||_1}{4},
}
\\
\\
{\displaystyle
||\mathcal{E}||_1\equiv \max_\rho ||\mathcal{E}(\rho)||_1,
}
\end{array}
\label{eq:tracenorm}
\end{equation}
where $||X||_1\equiv {\rm Tr} \sqrt{X^\dagger X}$ denotes the 
trace norm. 

The superoperator trace distance 
$||\mathcal{E}_1-\mathcal{E}_2||_1$
is, however, not a good 
definition of the distance between two quantum operations.
The point is that in general it is possible to exploit quantum 
entanglement to increase the distinguishability of two quantum channels.
In this case, an ancillary Hilbert space $\mathcal{K}$ is introduced,
the input state $\xi$ is a density matrix in $\mathcal{K}\otimes\mathcal{H}$,
and the quantum operations are trivially extended to $\mathcal{K}$. 
That is to say, the output states to discriminate are 
$(I_{\mathcal K}\otimes \mathcal{E}_1)\xi$
and $(I_{\mathcal K}\otimes \mathcal{E}_2)\xi$,
where $\mathcal{I}_\mathcal{K}$ is the identity map acting on
$\mathcal{K}$. The minimal error probability reads 
\begin{equation}
\begin{array}{c}
{\displaystyle
p_E= \frac{1}{2}-\frac{||\mathcal{E}_1-\mathcal{E}_2||_\diamond}{4},
}
\\
\\
{\displaystyle
||\mathcal{E}||_\diamond \equiv 
\max_\xi ||(\mathcal{I}_{\mathcal K}\otimes \mathcal{E})\xi||_1,
}
\end{array}
\label{eq:diamondnorm}
\end{equation}
where $||\mathcal{E}||_\diamond$ denotes the diamond norm 
of $\mathcal{E}$. 
It is clear from definition (\ref{eq:diamondnorm})
that 
\begin{equation}
||\mathcal{E}||_\diamond=||\mathcal{I}_{\mathcal{K}}\otimes \mathcal{E}||_1
\ge ||\mathcal{E}||_1
\end{equation}
and therefore $p_E\le p_E'$, so that it can be convenient to use an ancillary system
to better discriminate two quantum operations after a single channel use. 
The two quantum channels $\mathcal{E}_1$ and $\mathcal{E}_2$ become perfectly
distinguishable ($p_E=0$) when their diamond distance 
$||\mathcal{E}_1-\mathcal{E}_2||_\diamond=2$.

It turns out that the diamond norm does not depend on $\mathcal{K}$, provided
${\rm dim}(\mathcal{K})\ge {\rm dim}(\mathcal{H})$~\cite{kitaev}.
Due to the convexity of the trace norm, it can be shown that the 
maximum in both Eqs. (\ref{eq:tracenorm}) and (\ref{eq:diamondnorm}) is achieved
for pure input states~\cite{RW05}. 

\subsection{Kitaev's characterization of the diamond norm}

Kitaev provided a different equivalent characterization of the 
diamond norm, see, e.g.,~\cite{kitaev,watrous}. Any
superoperator (not necessarily completely positive) 
$\mathcal{E}: L(\mathcal{H})\to L(\mathcal{H})$, with $L(\mathcal{H})$
space of linear operators mapping $\mathcal{H}$ to itself, can be 
expressed as
\begin{equation}
\mathcal{E}(X)={\rm Tr}_{\mathcal R} (A X B^\dagger),
\label{eq:EX}
\end{equation}
where $X\in L(\mathcal{H})$, $A$ and $B$ linear operators from 
$\mathcal{H}$ to $\mathcal{R}\otimes \mathcal{H}$, 
with $\mathcal{R}$ auxiliary Hilbert space 
and ${\rm dim}(\mathcal{R})\le [{\rm dim}(\mathcal{H})]$.  
It is then possible to define completely positive superoperators
$\Psi_A, \Psi_B:L({\mathcal H})\to L({\mathcal R})$:
\begin{equation}
\Psi_A(X)={\rm Tr}_{\mathcal H}(A X A^\dagger),
\quad
\Psi_B(X)={\rm Tr}_{\mathcal H}(B X B^\dagger).
\label{eq:PsiAPsiB}
\end{equation}
Note that the space ${\mathcal H}$ is traced out in the 
definition of $\Psi_A$, $\Psi_B$, rather than the space
${\mathcal R}$. 
Finally, it turns out that~\cite{kitaev,watrous} 
\begin{equation}
||\mathcal{E}||_\diamond =F_{\rm max}(\Psi_A,\Psi_B),
\end{equation}
where $F_{\rm max}(\Psi_A,\Psi_B)$ is the maximum output fidelity of 
$\Psi_A$ and $\Psi_B$, defined as 
\begin{equation}
F_{\rm max}(\Psi_A,\Psi_B)=
\max_{\rho_1,\rho_2} F[\Psi_A(\rho_1),\Psi_B(\rho_2)],
\label{eq:dnormkitaev}
\end{equation}
where $\rho_1,\rho_2$ are density matrices in $\mathcal{H}$,
and the fidelity $F$ is defined as
\begin{equation}
F(\Psi_A,\Psi_B)={\rm Tr} \sqrt{\Psi_A^{1/2}\Psi_B\Psi_A^{1/2}}.
\label{eq:kitaevoptimization}
\end{equation}
Note that $\Psi_A$, $\Psi_B$ are not density matrices: 
the conditions ${\rm Tr}(\Psi_A)=1$, ${\rm Tr}(\Psi_B)=1$
are not satisfied. 

\section{Computing the diamond norm}
\label{sec:methods}

We numerically compute the distance (induced by the diamond norm) 
between two quantum channels $\mathcal{E}_1$ and $\mathcal{E}_2$
using two Monte-Carlo algorithms. 
The first one is based on the direct computation of 
$||\mathcal{E}_1-\mathcal{E}_2||_\diamond$, with the output states 
$(\mathcal{I}_{\mathcal K}\otimes \mathcal{E}_1)\xi$
and $(\mathcal{I}_{\mathcal K}\otimes \mathcal{E}_2)\xi$
in Eq.~(\ref{eq:diamondnorm})
computed from $\xi$ taking advantage of the Fano representation 
of quantum states and quantum operations. The second Monte-Carlo
algorithm uses the Kitaev's representation of the diamond norm 
to compute the maximum output fidelity $F_{\rm max}$ 
of Eq.~(\ref{eq:dnormkitaev}). In the following, we will refer
to the two above Monte-Carlo algorithms as F-algorithm and K-algorithm,
respectively. For the sake of simplicity 
we will confine ourselves to one-qubit quantum 
channels, even though the two algorithms can be easily formulated for
two- or many-qubit channels.   

\subsection{The F-algorithm}
\label{subsec:Fmethod}

In this section we describe the 
F-algorithm, which we will use to directly compute the diamond norm
(\ref{eq:diamondnorm}), with the maximum taken over a large number
of randomly chosen input states $\xi$. 
A convexity argument shows that it is sufficient to optimize
over pure input states $\xi=|\Psi\rangle\langle\Psi|$~\cite{RW05}. 
For one-qubit channels, it is enough to add a single ancillary
qubit when computing the diamond norm~\cite{kitaev}. 
Therefore, we can write
\begin{equation}
|\Psi\rangle=C_{00}|00\rangle+C_{01}|01\rangle+C_{10}|10\rangle
+C_{11}|11\rangle,
\end{equation}
with 
\begin{equation}
\begin{array}{l}
{\displaystyle
C_{00}=\cos\theta_1 \cos \theta_2,
}
\\
{
C_{01}=\cos\theta_1 \sin \theta_2 e^{i\phi_1},
}
\\
{
C_{10}=\sin\theta_1 \cos \theta_3 e^{i\phi_2},
}
\\
{
C_{11}=\sin\theta_1 \sin \theta_3 e^{i\phi_3},
}
\end{array}
\label{eq:thetaphi}
\end{equation}
where the angles $\theta_i\in\left [0,\frac{\pi}{2}\right]$ and the 
phase $\phi_i\in [0,2\pi]$. Hence, the maximization in the diamond 
norm is over the 6 real parameters $\theta_1,\theta_2,\theta_3$,
and $\phi_1,\phi_2,\phi_3$. Of course, the number of parameters can 
be reduced for specific channels when there are symmetries. 

Let $\mathcal{E}_1$ and $\mathcal{E}_2$ denote the two single-qubit
superoperators we would like to distinguish. 
The output states $\xi_1^\prime\equiv 
(\mathcal{I}_{\mathcal K}\otimes \mathcal{E}_1)\xi$ and 
$\xi_2^\prime\equiv 
(\mathcal{I}_{\mathcal K}\otimes \mathcal{E}_2)\xi$
can be conveniently computed using the Fano representation. 
Any two-qubit state can be written in the Fano form as 
follows~\cite{fano,eberly,mahler}:
\begin{equation}
\xi=\frac{1}{4} \sum_{\alpha,\beta=x,y,z,I}
R_{\alpha\beta}
\sigma_{\alpha}\otimes\sigma_{\beta},
\label{eq:fanoform}
\end{equation}
where $\sigma_x$, $\sigma_y$, and $\sigma_z$ are the Pauli matrices,
$\sigma_I\equiv \openone$, and
\begin{equation}
R_{\alpha\beta}=
{\rm Tr}[(\sigma_{\alpha}\otimes \sigma_{\beta}) \xi].
\label{eq:Ralphabeta}
\end{equation}
Note that the normalization condition ${\rm Tr}(\xi)=1$ implies
$R_{II}=1$.
Moreover, the coefficients $R_{\alpha\beta}$ are real due to 
the hermicity of $\xi$.
Eqs.~(\ref{eq:fanoform}) and (\ref{eq:Ralphabeta}) allow us to go from
the standard representation for the density matrix 
(in the $\{|0\rangle\equiv |00\rangle,
|1\rangle\equiv |01\rangle,|2\rangle\equiv|10\rangle,
|3\rangle\equiv |11\rangle\}$ basis) 
to the Fano representation, and vice versa. 
It is convenient to write the coefficients  
$R_{\alpha\beta}$ as a column vector,
\begin{equation}
\begin{array}{c}
{\bf R}=
[R_{xx},
R_{xy},
R_{xz},
R_{xI},
R_{yx},
R_{yy},
\\
\\
R_{yz},
R_{yI},
R_{zx},
R_{zy},
R_{zz},
R_{zI},
R_{Ix},
R_{Iy},
R_{Iz},
R_{II}]^T.
\end{array}
\end{equation}
Then the quantum operations 
$\mathcal{I}_{\mathcal K}\otimes \mathcal{E}_1$
and $\mathcal{I}_{\mathcal K}\otimes \mathcal{E}_2$
map, in the Fano representation, ${\bf R}$ into
${\bf R_1^\prime}=\mathcal{M}_1^{(2)} {\bf R}$  
and ${\bf R_2^\prime}=\mathcal{M}_2^{(2)} {\bf R}$,
respectively,
where $\mathcal{M}_1^{(2)}$ and $\mathcal{M}_2^{(2)}$ 
are affine transformation matrices.
Such matrices have a simple block structure:
\begin{equation}
\mathcal{M}_i^{(2)}=\mathcal{I}^{(1)}\otimes \mathcal{M}_i^{(1)},
\end{equation}
with $\mathcal{I}^{(1)}$ and $\mathcal{M}_i^{(1)}$ 
$4\times 4$ affine transformation matrices corresponding to the quantum 
operations $\mathcal{I}_{\mathcal K}$ and $\mathcal{E}_i$
(of course, $\mathcal{I}^{(1)}$ is the identity matrix).  
Matrices $\mathcal{M}_i^{(1)}$ directly determines the 
transformation, induced by $\mathcal{E}_i$, of the 
single-qubit Bloch-sphere coordinates
$(x,y,x)$. Given the Fano
representation for a single qubit,
$\rho=\frac{1}{2}\sum_{\alpha} r_\alpha \sigma_\alpha$,
then ${\bf r}=[x,y,z,1]^T$ and 
${\bf r_i^\prime}\equiv\mathcal{M}_i^{(1)} {\bf r}$.
We point out that, while one could compute 
$\xi_i^\prime$ 
from the Kraus representation of the superoperator
$\mathcal{I}_{\mathcal K}\otimes \mathcal{E}_i$,
the advantage of the Fano representation is that the matrix 
elements of $\mathcal{M}_i$ are directly 
related to the transformation of the expectation values of the system's
polarization measurements~\cite{qcbook,BFS,BS09}.

Finally, we compute the trace distance between 
$\xi_1^\prime$ and $\xi_2^\prime$ as
\begin{equation}
||\xi_1^\prime-\xi_2^\prime||_1=
\sum_k |\lambda_k|,
\label{eq:tracenormxi}
\end{equation}
where $\lambda_1,...,\lambda_4$ are the eigenvalues of
$\xi_1^\prime-\xi_2^\prime$.

\subsection{The K-algorithm}
\label{subsec:Kmethod}

We consider a special and unnormalized state in the extended Hilbert space 
$\mathcal{K}\otimes\mathcal{H}$:
\begin{equation}
|\alpha\rangle=\sum_j |j_{\mathcal K}\rangle|j_{\mathcal H}\rangle,
\end{equation}
where ${\rm dim}(\mathcal{K})={\rm dim}(\mathcal{H})$ and
$\{|j_{\mathcal K}\rangle\}$, $\{|j_{\mathcal H}\rangle\}$
are orthonormal bases for $\mathcal{K}$, $\mathcal{H}$.
The state $|\alpha\rangle$ is, up to a normalization factor, 
a maximally entangled state in $\mathcal{K}\otimes \mathcal{H}$. 

We define an operator $\sigma$ on $\mathcal{K}\otimes \mathcal{H}$:
\begin{equation}
\sigma=(\mathcal{I}_{\mathcal K}\otimes \mathcal{E}) 
(|\alpha\rangle\langle\alpha|),
\end{equation}
where $\mathcal{E}=\mathcal{E}_1-\mathcal{E}_2$ is the difference of two 
quantum operations but is not a quantum operation itself. 
That is, $\mathcal{E}$ is linear but not trace preserving 
and completely positive. 

Using the singular value decomposition of $\sigma$, we can write
\begin{equation}
\sigma=\sum_{i=1}^M |u^{(i)}\rangle \langle v^{(i)}|,
\end{equation}
where $M$ is the rank of $\sigma$ and $|u^{(i)}\rangle$, $|v^{(i)}\rangle$
are vectors in $\mathcal{K}\otimes \mathcal{H}$. 
The operator $\sigma$ completely specifies $\mathcal{E}$ and 
can be exploited to express $\mathcal{E}$ as 
\begin{equation}
\mathcal{E}(X)=\sum_{i=1}^M A^{(i)} X B^{(i)\dagger}.
\label{eq:AiBisum}
\end{equation}

The operators $A^{(i)}$ and $B^{(i)}$ can be derived by generalizing 
the construction of the operator-sum representation for quantum 
operations (see, for instance, Sec.~8.2.4 in Ref.~\cite{nielsen}).
Given a generic state $|\psi\rangle=\sum_j \psi_j |j_{\mathcal H}\rangle$
in $\mathcal{H}$, we define a corresponding state in $\mathcal{K}$:
$|\tilde{\psi}\rangle=\sum_j \psi_j^\star |j_{\mathcal K}\rangle$.
Next, we define 
\begin{equation}
A^{(i)}=\langle \tilde{\psi} | u^{(i)}\rangle.
\end{equation}
For instance, in the single-qubit case
$|u^{(i)}\rangle=[u^{(i)}_1,u^{(i)}_2,u^{(i)}_3,u^{(i)}_4]^T$
and we obtain
\begin{equation}
A^{(i)}=
\left[
\begin{array}{cc} 
u^{(i)}_1 & u^{(i)}_3\\
u^{(i)}_2 & u^{(i)}_4
\end{array}
\right].
\end{equation}
Similarly, we define 
\begin{equation}
B^{(i)}=\langle \tilde{\psi} | v^{(i)}\rangle.
\end{equation}
Finally, it can be checked that with the above defined 
operators $A^{(i)}$, $B^{(i)}$ we can express $\mathcal{E}$
by means of Eq.~(\ref{eq:AiBisum}).

We can now give explicit expressions for $\Psi_A$ and  $\Psi_B$ in
Eq.~(\ref{eq:PsiAPsiB}):
\begin{equation}
[\Psi_A(X)]_{ij}=\sum_{\alpha,m,n} A^{(i)}_{\alpha m} X_{mn}
A^{(j)\star}_{\alpha n},
\end{equation}
\begin{equation}
[\Psi_B(X)]_{ij}=\sum_{\alpha,m,n} B^{(i)}_{\alpha m} X_{mn}
B^{(j)\star}_{\alpha n},
\end{equation}
with $1\le i,j \le M$. Therefore,  
$\Psi_A$ and $\Psi_B$ are $M\times M$ matrices. 
In the single-qubit case, $M\le 4$ and  
to compute the diamond norm through Eq.~(\ref{eq:dnormkitaev})
we need to calculate eigenvalues and eigenvectors of
matrices of size $M$. 
A simpler but less efficient decomposition of $\mathcal{E}$,
$\Psi_A$, $\Psi_B$ is provided in Appendix~\ref{app:Kraus2}.

For single-qubit channels the optimization 
(\ref{eq:dnormkitaev}) is over 6 real parameters, 
3 for $\rho_1$ and 3 for $\rho_2$ (for instance, the Bloch-sphere
coordinates of $\rho_1$ and $\rho_2$). 
As discussed in Sec.~\ref{subsec:Kmethod},
the same number of parameters 
are needed in the F-method. 
However, the F-method has computational advantages in that 
only the eigenvalues of the $4\times 4$ matrix 
$\xi_1^\prime- \xi_2^\prime$ are required, while in the K-method
we need to evaluate both eigenvalues and eigenvectors 
of matrices in general of the same size ($M=4$). 
Moreover, the singular-values decomposition of matrix $\sigma$ 
is needed.
Besides computational advantages, the F-method is physically
more transparent, as it is based on affine maps, whose matrix  
elements have physical meaning, being directly
related to the transformation of the expectation values of the system's
polarization measurements~\cite{qcbook,BFS,BS09}.

\section{Examples for single-qubit quantum channels}
\label{sec:1qchannels}

In this section, we illustrate the working of the F- and K-methods 
for the case, most significant for present-day implementations, of 
single-qubit quantum channels.

\subsection{Pauli channels}
\label{sec:Pauli}

We start by considering the case of Pauli channels,
\begin{equation}
\mathcal{E}_i(\rho) = \sum_{\alpha=x,y,z,I} (q_\alpha)_i 
\sigma_\alpha \rho \sigma_\alpha,\;\;\;\sum_\alpha (q_\alpha)_i=1,
\;\; i=1,2,
\label{eq:Pauli}
\end{equation}
for which the diamond norm can be evaluated analytically~\cite{sacchi}:
\begin{equation}
||\mathcal{E}||_\diamond=||\mathcal{E}_1-\mathcal{E}_2||_\diamond=
\sum_\alpha |(q_\alpha)_1-(q_\alpha)_2|,
\label{eq:dPauli}
\end{equation}
this value of the diamond norm being achieved for maximally entangled 
input states. The Pauli-channel case will serve as a testing ground 
for the F- and K-algorithms and help us develop a physical
and geometrical intuition. 

\begin{figure}[t]
\centering
\includegraphics[width=7.0cm]{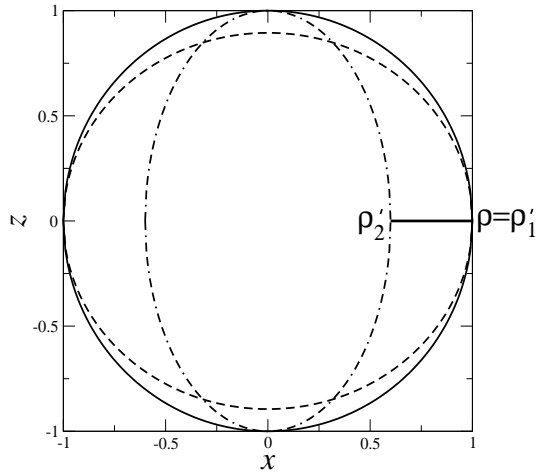}
\caption{Schematic drawing of the trace (and diamond) distance distance
between the bit-flip and the phase-flip channels.
\label{fig:bitphaseflip}}
\end{figure}

Let us focus on a couple of significant examples. We first consider
the bit-flip and the phase-flip channels:
\begin{equation}
\begin{array}{l}
{\displaystyle
\mathcal{E}_1(\rho)=\frac{1+c_1}{2}\rho+
\frac{1-c_1}{2}\sigma_x\rho\sigma_x,
}
\\
\\
{\displaystyle
\mathcal{E}_2(\rho)=\frac{1+c_2}{2}\rho+
\frac{1-c_2}{2}\sigma_z\rho\sigma_z,
}
\end{array}
\end{equation}
with $0\le c_1,c_2\le 1$.
We then readily obtain from Eq.~(\ref{eq:dPauli}) that
\begin{equation}
||\mathcal{E}||_\diamond=||\mathcal{E}_1-\mathcal{E}_2||_\diamond=
\max\{1-c_1,1-c_2\}.
\end{equation}
For this example, the diamond norm coincides with the trace norm,
$||\mathcal{E}||_\diamond=||\mathcal{E}||_1$, 
and therefore has a simple geometrical interpretation: For
a single qubit the trace distance between two single-qubit states 
is equal to the Euclidean distance between them on the Bloch 
ball~\cite{nielsen}. Superoperators $\mathcal{E}_1$ 
and $\mathcal{E}_2$ map the Bloch sphere into an ellipsoid 
with $x$ (for the bit-flip channel) and $z$ (for the phase-flip
channel) as symmetry axis. If we call $(x,y,z)$, 
$(x_1',y_1',z_1')$, and
$(x_2',y_2',z_2')$,
the initial Bloch-sphere coordinates and the new coordinates 
after application of quantum operations $\mathcal{E}_1$ and 
$\mathcal{E}_2$, respectively, we obtain
\begin{equation}
\begin{array}{l}
x_1'=x, \;\; y_1'=c_1 y, \;\; z_1'=c_1 z,
\\
\\
x_2'=c_2 x, \;\; y_2'=c_2 y, \;\; z_2'=z.
\end{array}
\end{equation}
The geometrical meaning of the trace norm for the present example 
is clear from Fig.~\ref{fig:bitphaseflip}: the length of the 
line segment $\overline{\rho\rho_2'}=\overline{\rho_1'\rho_2'}$ 
is the trace (and the diamond) distance
$||\mathcal{E}_1-\mathcal{E}_2||_1=
||\mathcal{E}_1-\mathcal{E}_2||_\diamond$ 
(note that in this figure 
$1-c_2>1-c_1$).

\begin{figure}[t]
\centering
\includegraphics[width=7.0cm]{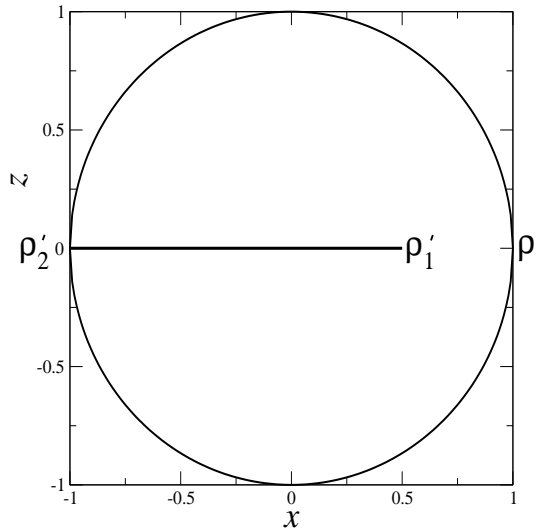}
\caption{Schematic drawing of the trace distance distance
between the two Pauli channels of Eq.~(\ref{eq:sacchi}).
\label{fig:sacchi}}
\end{figure}

As a further example, we discuss a special instance of the 
channels considered in Ref.~\cite{sacchi}:
\begin{equation}
\begin{array}{l}
{\displaystyle
\mathcal{E}_1(\rho)=\frac{1}{2}\rho+
\frac{1}{4}\sigma_x\rho\sigma_x+
\frac{1}{4}\sigma_y\rho\sigma_y,
}
\\
\\
{\displaystyle
\mathcal{E}_2(\rho)=
\sigma_z\rho\sigma_z.
}
\end{array}
\label{eq:sacchi}
\end{equation}
In this case, 
\begin{equation}
\begin{array}{l}
{\displaystyle
x_1'=\frac{1}{2}x, \;\; y_1'=\frac{1}{2} y, \;\; z_1'=0,
}
\\
\\
x_2'=-x, \;\; y_2'=-y, \;\; z_2'=z.
\end{array}
\end{equation}
The trace norm $||\mathcal{E}||_1=\frac{3}{2}$,
as shown in Fig.~\ref{fig:sacchi}, where 
$||\mathcal{E}||_1$ is given by the length of the segment 
$\overline{\rho_1'\rho_2'}$. 
On the other hand, the two channels $\mathcal{E}_1$ and 
$\mathcal{E}_2$ are perfectly distinguishable, as
we readily obtain from Eq.~(\ref{eq:dPauli}) that 
$||\mathcal{E}||_\diamond=2$, this value being achieved
by means of maximally entangled input states.
Therefore, in this example 
$||\mathcal{E}||_\diamond>||\mathcal{E}||_1$, that is,
entangled input states improve the distinguishability of the 
two channels. 

\begin{figure}[t]
\centering
\includegraphics[width=8.0cm]{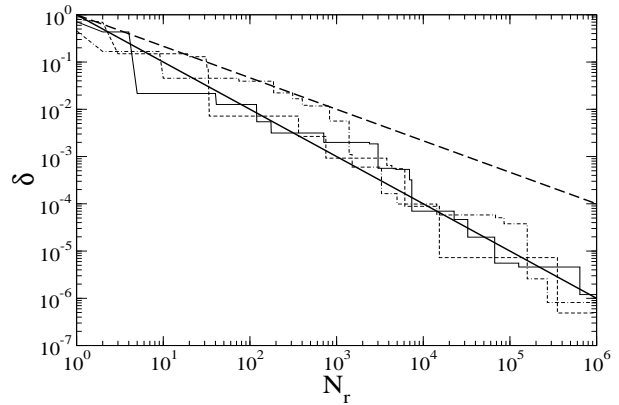}
\caption{Error $\delta$ in evaluating the diamond distance
between channels (\ref{eq:sacchi}) by means of
the F-algorithm after $N_r$ randomly chosen initial conditions.
Three different runs are shown, together with the
$1/N_r$ (full line) and $(1/N_r)^{2/3}$ (dashed line) 
dependences.
\label{fig:test}}
\end{figure}

Channels (\ref{eq:sacchi}) are very convenient to illustrate the 
convergence properties of the F- and K-algorithms. 
Let us first consider the F-algorithm.
If the maximum of $||\xi_1^\prime-\xi_2^\prime||_1$, with 
$\xi_i^\prime=
(\mathcal{I}_{\mathcal K}\otimes \mathcal{E}_i)\xi$ 
($i=1,2$)
is taken over $N_r$ random initial conditions, 
then the obtained value differs from the diamond 
norm $||\mathcal{E}||_\diamond$ by an amount 
$\delta(N_r)$ which must converge to zero when 
$N_r\to\infty$. 
To get convergence to $||\mathcal{E}||_\diamond$ 
for channels (\ref{eq:sacchi}) it is 
enough to optimize over real initial conditions.
Numerical results, shown in Fig.~\ref{fig:test} for a few runs,
are consistent with $\delta (N_r) \sim 1/N_r$. A rough argument
can be used to explain the $1/N_r$ dependence. Assuming a smooth
quadratic dependence of the distance
$D(\xi)\equiv ||\xi_1^\prime - \xi_2^\prime||_1$
on the parameters for optimization
[the angles $\theta_i$ in (\ref{eq:thetaphi}),
with $\theta_1\in \left[0,\frac{\pi}{2}\right]$,
$\theta_2,\theta_2\in[0,2\pi)$]
for $\xi$ around the value $\xi_0$ optimizing $D$,
then $|D(\xi)-D(\xi_0)| \sim \delta$ when the distance
between $\xi$ and $\xi_0$ is of the order of $\sqrt{\delta}$.
The number $N_r$ of randomly distributed initial conditions typically
requested to get a point satisfying $||\xi-\xi_0||_1<\sqrt{\delta}$ is
of the order of $(1/\sqrt{\delta})^{n_p}$, where $n_p$ is the number
of parameters for optimization. Therefore,
$\delta(N_r)\sim (1/N_r)^{2/\delta}$. In the example of
Fig.~\ref{fig:test}, $n_p=3$ leading to
$\delta(N_r)\sim (1/N_r)^{2/3}$. On the other hand,
numerical data exhibit a $1/N_r$ dependence.
This fact has a simple explanation: the maximum is achieved
for Bell states, which are invariant under rotations. 
Hence, the maximum distance is obtained
not on a single point but on a curve and the number 
$N_r$ of initial conditions requested to get a point
within a trace distance smaller than $\sqrt{\delta}$
from this curve is 
of the order of $(1/\sqrt{\delta})^{n_p-1}$, thus leading to
$\delta(N_r)\sim 1/N_r$.

With regard to the K-algorithm,
we have observed in the example of Eq.~(\ref{eq:sacchi}) the 
same $1/N_r$ convergence to the expected asymptotic 
value $||\mathcal{E}||_\diamond=2$. However, the cost per 
initial condition in the K-algorithm is much larger than
in the F-algorithm, in agreement with the general discussion 
of Sec.~\ref{subsec:Kmethod}.
Furthermore, the physical meaning of the F-method is much more 
transparent. For instance, for the Pauli channels 
(\ref{eq:Pauli}) the affine transformation matrix 
$\mathcal{M}_i^{(1)}$ such 
that ${\bf R_i^\prime}=(\mathcal{I}^{(1)}\otimes 
\mathcal{M}_i^{(1)}) {\bf R}$
has a simple diagonal structure:
\begin{equation}
\mathcal{M}_i^{(1)}={\rm diag}
[(c_x)_i,(c_y)_i,(c_z)_i,1],
\end{equation}
where 
\begin{equation}
\begin{array}{l}
(c_x)_i=1-2[(q_y)_i+(q_z)_i],
\\
\\
(c_y)_i=1-2[(q_z)_i+(q_x)_i],
\\
\\
(c_z)_i=1-2[(q_x)_i+(q_y)_i].
\end{array}
\end{equation}
Matrix $\mathcal{M}_i^{(1)}$ simply accounts for the transformation, induced 
by $\mathcal{E}_i$, of the polarization measurements for the system 
qubit: $\alpha_i^\prime=(c_\alpha)_i \alpha$, with $\alpha=x,y,z$
Bloch-sphere coordinates.   

We have checked the computational advantages of the F-method 
also for all the other examples discussed in this paper. 
For this reason in what follows we shall focus on this method only.

\subsection{Nonunital channels}
\label{sec:nonunital}

In this section, we consider nonunital channels, that is, channels
that do not preserve the identity. Therefore, in  
contrast to the Pauli channels considered in Sec.~\ref{sec:Pauli},
single-qubit nonunital channels displace the center of the 
Bloch sphere. Such channel are physically relevant in the 
description, for instance, of energy dissipation in open quantum systems, 
the simplest case being the amplitude damping 
channel~\cite{nielsen,qcbook}. 

As a first illustrative example, we compute the distance between
two channels $\mathcal{E}_1$ and $\mathcal{E}_2$
corresponding to displacements of the Bloch sphere
along the $+x$- and $+z$-direction~\cite{BFS}. 
In the Fano representation the affine
transformation matrices $\mathcal{M}_i^{(1)}$
corresponding to maps $\mathcal{E}_i$ have a simple structure:
\begin{equation}
\mathcal{M}_1^{(1)}=
\left[
\begin{array}{cccc}
C_x^2 & 0 & 0 & S_x^2 \\
0     & C_x & 0 & 0 \\
0     & 0  & C_x & 0 \\
0     & 0  & 0 & 1
\end{array}
\right],
\end{equation}
\begin{equation}
\mathcal{M}_2^{(1)}=
\left[
\begin{array}{cccc}
C_z  & 0 & 0 & 0 \\
0     & C_z & 0 & 0 \\
0     & 0  & C_z^2 & S_z^2 \\
0     & 0  & 0 & 1
\end{array}
\right],
\end{equation}
where we have used the shorthand notation $C_x\equiv\cos\theta_x$ 
and $C_z\equiv\cos\theta_z$. The two channels depend parametrically
on $\theta_x,\theta_z\in \left[0,\frac{\pi}{2}\right]$. The 
limiting cases $\theta_x=0$ and $\theta_z=0$  
correspond to $\mathcal{E}_1=\mathcal{I}$ and 
$\mathcal{E}_2=\mathcal{I}$, respectively.
Moreover, for
$\theta_x=\frac{\pi}{2}$ quantum operation 
$\mathcal{E}_1$ maps 
the Bloch ball onto the single point
$(x_1'=1,y_1'=0,z_1'=0)$; for  
$\theta_z=\frac{\pi}{2}$ the mapping operated 
by $\mathcal{E}_2$ is onto the north pole of the Bloch sphere,
$(x_2'=0,y_2'=0,z_2'=1)$. 

\begin{figure}[t]
\centering
\includegraphics[width=8.0cm]{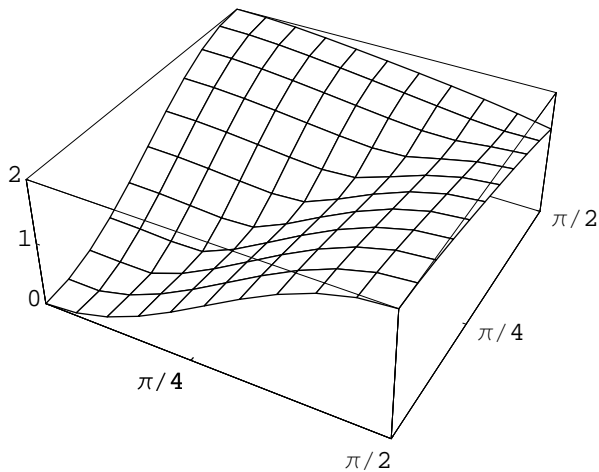}
\caption{Trace norm 
$||\mathcal{E}||_1=||\mathcal{E}_1-\mathcal{E}_2||_1$, 
where $\mathcal{E}_1$ and $\mathcal{E}_2$ 
are displacement channels along the $+x$- and $+z$-axis
of the Bloch sphere. 
\label{fig:dispxz}}
\end{figure}

The numerically computed trace distance
$||\mathcal{E}_1-\mathcal{E}_2||_1$ 
is shown in Fig.~\ref{fig:dispxz}, as a function of the 
parameters $\theta_x$ and $\theta_z$. We gathered numerical
evidence that for such channels the diamond distance 
equals the trace distance.

\begin{figure}[t]
\centering
\includegraphics[width=8.0cm]{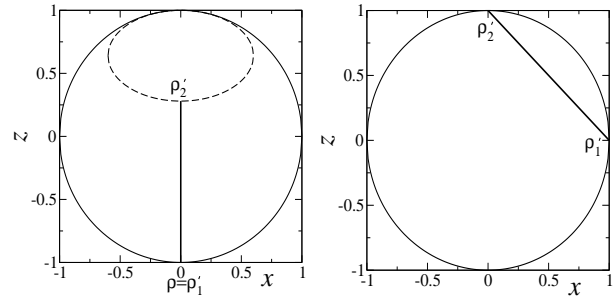}
\caption{Bloch-sphere visualization of the trace norm 
$||\mathcal{E}||_1$ in the limiting cases 
$\theta_x=0$ (left) and $\theta_x=\theta_z=\frac{\pi}{2}$
(right).
\label{fig:blochdispxz}}
\end{figure}

It is interesting to examine the analytical solutions for
two limiting cases: (i) $\theta_x=0$ 
(the case $\theta_z=0$ is analogous) and
(ii) $\theta_x=\theta_z=\frac{\pi}{2}$.
In the first case, 
$\mathcal{E}_1$ maps the Bloch sphere into an ellipsoid
with $z$ as symmetry axis~\cite{BFS}, while 
$\mathcal{E}_2=\mathcal{I}$. The trace norm is given 
by the length of the line segment 
$\overline{\rho\rho_2'}=
\overline{\rho_1'\rho_2'}$ shown in Fig.~\ref{fig:blochdispxz} (left),
\begin{equation}
||\mathcal{E}||_1=2\sin^2\theta_z.
\label{eq:tnormdispz}
\end{equation}

\begin{figure}[t]
\centering
\includegraphics[width=8.0cm]{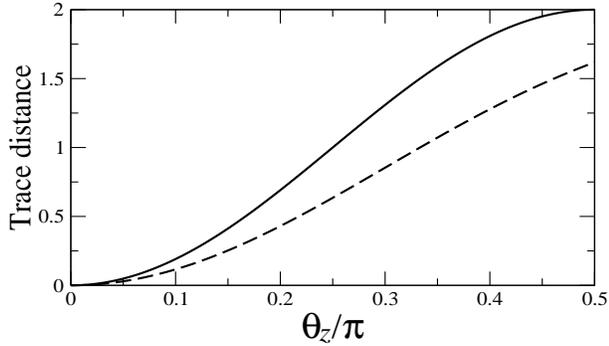}
\caption{Comparison between 
$||\mathcal{E}_1-\mathcal{E}_2||_1$ (upper curve) and 
$||\xi_1^\prime-\xi_2^\prime||_1$ (lower curve)
for $\theta_x=\frac{\pi}{2}$.
\label{fig:checknorm}}
\end{figure}

It is interesting to remark that, in contrast to the Pauli channels,
the optimal input state is not a maximally entangled input state. 
In the limiting case (i) the channels are strictly better
discriminated by means of an appropriate separable input state,
i.e., the south pole of the Bloch sphere
(see the left plot of Fig.~\ref{fig:blochdispxz})) rather than by 
Bell states. Indeed, given the maximally entangled input 
state
\begin{equation}
\xi=|\Psi\rangle\langle\Psi|,
\quad
|\Psi\rangle=\frac{1}{\sqrt{2}}\,(|00\rangle+|11\rangle),
\end{equation}
we obtain 
\begin{equation}
\xi_1^\prime-\xi_2^\prime=
\frac{1}{2}\,\left[
\begin{array}{cccc}
0 & 0 & 0 & C_z-1 \\
0 & 0 & 0 & 0 \\
0 & 0 & S_z^2 & 0 \\
C_z-1 & 0 & 0 & -S_z^2
\end{array}
\right].
\end{equation}  
The trace distance between $\xi_1^\prime$ and $\xi_2^\prime$ 
is then computed by means of Eq.~(\ref{eq:tracenormxi}):
\begin{equation}
\begin{array}{c}
{\displaystyle
||\xi_1^\prime-\xi_2^\prime||_1=
\frac{S_z^2}{2}+
\,\frac{\left|S_z^2+\sqrt{S_z^4+4(1-C_z)^2}\right|}{4}
}
\\
\\
{\displaystyle
+\frac{\left|S_z^2-\sqrt{S_z^4+4(1-C_z)^2}\right|}{4}.
}
\end{array}
\label{eq:dnormdispz}
\end{equation}
As shown in Fig.~\ref{fig:checknorm}, 
$||\xi_1^\prime-\xi_2^\prime||_1<||\mathcal{E}||_1$ for
any $C_z>0$. 

In case (ii), for any initial state $\rho$, 
the Bloch coordinates of $\rho_1^\prime$ and $\rho_2^\prime$ are 
given by $(1,0,0)$ and $(0,0,1)$. The trace norm is given by 
the distance between these two points, $||\mathcal{E}||_1=\sqrt{2}$
[see Fig.~\ref{fig:blochdispxz} (right)]. 
In this case, given an input Bell state or any other two-qubit
input state $\xi$, the trace distance $||\xi_1^\prime-\xi_2^\prime||_1
=\sqrt{2}=||\mathcal{E}||_1$. Thus, there is no advantage in using
an ancillary system.

As a further example, we compute the distance between the 
depolarizing channel $\mathcal{E}_1$ and the
nonunital channel $\mathcal{E}_2$ 
corresponding to the displacement along the $+z$-axis
of the Bloch sphere. The depolarizing channel belongs to
the class of Pauli channels (\ref{eq:Pauli}), with 
$q_I=1-p$ and $q_x=q_y=q_z=\frac{p}{3}$. This channel 
contracts the Bloch sphere by a factor $\left(1-\frac{4}{3}p\right)$,
with $0\le p \le \frac{3}{4}$.

\begin{figure}[t]
\centering
\includegraphics[width=8.0cm]{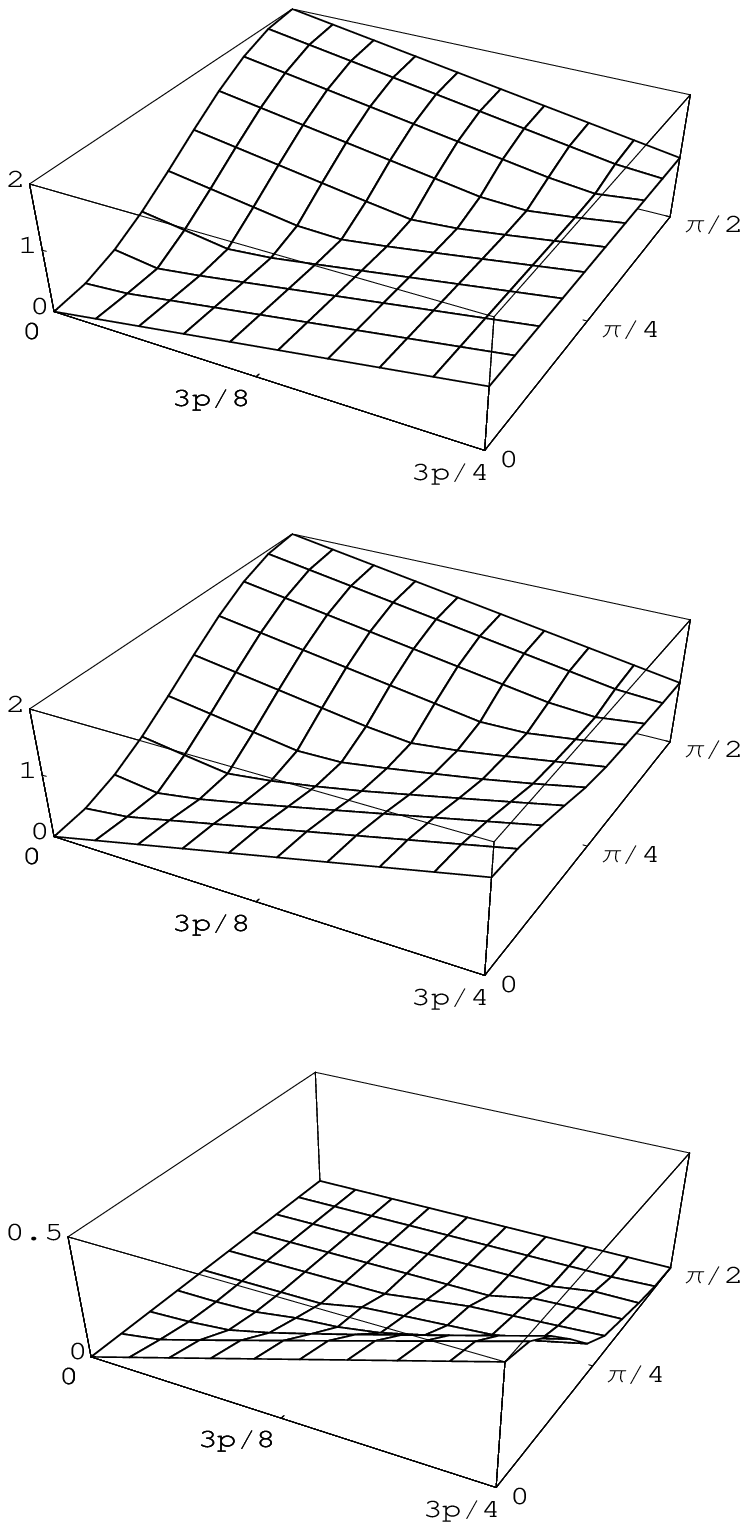}
\caption{From top to bottom: trace norm 
$||\mathcal{E}||_1=||\mathcal{E}_1-\mathcal{E}_2||_1$, 
diamond norm $||\mathcal{E}||_\diamond$, and 
their difference $||\mathcal{E}||_\diamond-||\mathcal{E}||_1$;
$\mathcal{E}_1$ is the depolarizing channel,
$\mathcal{E}_2$ the displacement along the $+z$-axis of
the Bloch sphere.
\label{fig:dispdep}}
\end{figure}

Fig.~\ref{fig:dispdep} shows the numerically computed 
$||\mathcal{E}||_1$ and $||\mathcal{E}||_\diamond$ as
well as their difference $||\mathcal{E}||_\diamond-
||\mathcal{E}||_1$. The use of an ancillary qubit
improves the distinguishability of the two channels. 
However, we remark again that 
maximally entangled input states can be 
detrimental. For instance, in the limiting case 
$p=0$, $\theta_z=\frac{\pi}{2}$ we obtain 
from Eqs.~(\ref{eq:tnormdispz}) and (\ref{eq:dnormdispz})
$||\mathcal{E}||_1=2>||\xi_1^\prime-\xi_2^\prime||_1
=\frac{1+\sqrt{5}}{2}$. A clear advantage is instead seen 
in another limiting case, $p=\frac{3}{4}$, $\theta_z=0$. 
The fully depolarizing channel maps each point of the Bloch 
ball onto its center, so that the trace distance 
is given by the radius of the Bloch sphere,
$||\mathcal{E}_1-\mathcal{E}_2||_1=1$. On the other hand,
by means of Eq.~(\ref{eq:dPauli}) we obtain 
$||\mathcal{E}_1-\mathcal{E}_2||_\diamond=\frac{3}{2}$.

\section{Conclusions}
\label{sec:conclusions}

We have shown that the distance between two quantum channels 
can be conveniently computed by means of a Monte-Carlo algorithm 
based on the Fano representation. The effectiveness of this algorithm 
is illustrated in the case, most relevant for present-day 
implementations of quantum information processing, of 
single-qubit quantum channels. A main computational advantage of 
this algorithms is that it is easily parallelizable. 
Furthermore, being based on the Fano representation, is 
enlights the physical meaning of the involved quantum channels: 
the matrix elements of the affine map representing a quantum channel
directly account for the evolution of
the expectation values of the system's
polarization measurements.
More generally, we believe that the Fano representation provides 
a computationally convenient and physically trasparent representation
of quantum noise.

\appendix

\section{Alternative decomposition of $\mathcal{E}$}
\label{app:Kraus2}

Given a superoperator $\mathcal{E}=\mathcal{E}_1-\mathcal{E}_2$,
with $\mathcal{E}_1$, $\mathcal{E}_2$ quantum operations,
we start from the Kraus representation~\cite{qcbook,nielsen} of 
$\mathcal{E}_1$ and $\mathcal{E}_2$:
\begin{equation}
\mathcal{E}_1(X)=\sum_{k=1}^{M_1} E_k X E_k^\dagger,
\quad
\mathcal{E}_2(X)=\sum_{k=1}^{M_2} F_k X F_k^\dagger,
\end{equation}
and define new operators:
\begin{equation}
\begin{array}{l}
{\displaystyle
\tilde{A}^{(2k-1)}=\frac{1}{\sqrt{2}}(F_k+G_k),
\quad
\tilde{A}^{(2k)}=\frac{1}{\sqrt{2}}(F_k-G_k),
}
\\
\\
{\displaystyle
\tilde{B}^{(2k-1)}=\tilde{A}^{(2k)},
\quad
\tilde{B}^{(2k)}=\tilde{A}^{(2k-1)},
}
\end{array}
\end{equation}
where $k=1,...,\overline{M}\equiv\max(M_1,M_2)$.
Note that, if $M_1>M_2$,
we set $G_k=0$ for $k=M_2+1,...,M_1$; vice-versa, if
$M_1<M_2$, $F_k=0$ for $k=M_1+1,...,M_2$.
It is easy to see that
\begin{equation}
\mathcal{E}(X)=\sum_{i=1}^{2 \overline{M}}
\tilde{A}^{(i)} X \tilde{B}^{(i)\dagger}.
\label{eq:tildeAB}
\end{equation}
In contrast to (\ref{eq:AiBisum}), the present decomposition 
of $\mathcal{E}(X)$ is simpler, in that no singular value decomposition
is required, but less efficient. Indeed, the maximum number of terms
in (\ref{eq:tildeAB}) is twice that of decomposition (\ref{eq:AiBisum}).   
This implies that, if $\Psi_A$, $\Psi_B$ are expressed in terms
of operators $\tilde{A}^{(i)}$, $\tilde{B}^{(i)}$ rather than 
$A^{(i)}$, $B^{(i)}$, to evaluate $F(\Psi_A,\Psi_B)$ we need to compute 
eigenvalues and eigenvectors of matrices of size $2\overline{M}$. 
In the single-qubit case, typically $2\overline{M}=8$.

\begin{acknowledgments}
We thank Massimiliano Sacchi for interesting comments on our work.
\end{acknowledgments}

\end{document}